\documentstyle[12pt]
{article}
\begin{document}

\textwidth 148mm
\textheight 225mm
\title {  Effect of Gluon Damping Rate
to the Viscosity Coefficient of the Quark-gluon Plasma at Finite
Chemical potential }

\author{Hou Defu $^*$ $\ \ \ \ \ $ Li Jiarong}

\baselineskip 24pt
\maketitle

\begin{center}
 {\small Institute of Particle Physics,
 Hua-Zhong Normal University\\
 Wuhan 430070, China\\
 E-mail: hzipp@ccnu.edu.cn\\ 
$* \ $ present address: Institute f\"ur theoretische physik,\\
93040 Uni Regensburg, 
Germany.\\
 E-mail: defu.hou@rphs1.physik.uni-regensburg.de\\}

\end{center}

\begin{abstract}
By considering the Debye screening and the damping rate of gluons,
the viscosity cofficient of the quark-gluon plasma was evaluated via 
 the real time finite temperature QCD in the relaxiation time approximation at finite temperature and
chemical potential. The results show that both 
the damping rate and quark chemical potential cause  considerable 
enhancements to  the viscosity coefficient of the hot  dense quark-gluon
plasma.

{\bf PCAC number(s): 12.38.Mh, 12.38.Bx}

\end{abstract}

\section{Introduction}

The calculation of the viscosity coeffcient in a  hot and dense system
is of interest both in the fields of high energy heavy ion collisions and
astrophysics [1-4].
The  dissipative processes in a quark-gluon plasma(QGP)
supposed to be formed
in ultrarelativstic heavy collision can be described by the
viscosity coefficient.

 Because of the estimated very short collision time of the ions
in the heavy ion collisions,
the equilibrium statistical approaches are probably inappropriate
for an adequate description of such a process, and the dissipative phenomena
are important, at least for the equilibration and the expansion phases of the
system. In principle,the dissipative processes in heavy ion collisions should
be described by non-equilibrium dynamical theory that is based on QCD.

There are two methods to calculate the shear viscosity in QGP. The first one
is by use of the relativstic kinetic theory. Starting from the Boltzmann
equation, the viscosity coefficient $\eta$ can be derived containing the
transport cross section[2,3]. The second one is from the kubo formulas[2].
 For example,  in the relaxation time approximation, the shear
viscosity coefficient was found to be $\eta=cT^3/\alpha_s^2\ln(1/\alpha_s)$,
the constant $c$ was estimated to be $1.02$,$0.28$,or $o.57$,whereas a
variational calculation gave $c=1.16$  with two quark flavors
in high temperature limit[2,3]. And Ref.3 also gave the result beyond the
leading logarithm approximation,
$$\eta={T^3\over \alpha_s^2}[ {0.11\over \ln({0.19\over \alpha_s})}
+ {0.37\over \ln({0.21\over \alpha_s})}].$$
On the other hand , basing on the kubo formula, Ref.2 gave
$\eta\le 2.6 T^3/\alpha_s$ and the lattice calculations obtained the value
near the phase transition as $0\le\eta/T^3\le 9.5$ [2].

It is known that the ensential nature of particles in a plasma at
finite temperature is that they no longer have a infinite lifetimes, but have
finite widths becaue of the collective effects. That is quite different from
the behavior at zero temperature[6,7]. In QGP Debye screening and damping of collective modes occur because of the interactions among particles and the heat bath.
Therefore it is reasonable  to take into account both the Debye screening and 
finite width effects when study the collective behaviors of the QGP,
, which might also influence the transport processes of QGP.
  
On the other hand, so far most all of the results of the shear viscosity cofficients in QGP are
calculated at finite temperature without taking into account of the baryon
density. Certainly by lattice simulation it is difficult to include the
baryon density 
,which is related to the baryo-chemical potential
$\mu_b$ through $\mu=\mu_b/3$ and measures the deviation from the balance of
quarks and and antiquarks.
However at RHIC energies,there might still exist a considerable amount of
stopping to $\mu/T\sim 1-2$[5] ,$\mu$
is the quark chemical potential.

In this paper we will investigate 
the damping rate effect of the transverse gluon on the shear
viscosity coefficient of QGP at both finite temperature and chemical potential
quark chemical potential. We will show explicitly both the finite width and
chemical potential cause considerable enhancements to the shear viscoisity.

The paper will be organized as follows. In Sec. II we will get the shear viscosity of quarks and gluons via real time QCD by considering both the Debye screening and finite width effects at finite chemical potential. In Sec.III we will
consider the running behavior of the coupling constant ane present numerical 
results. Finally we will summarize our results in Sec.IV. 
\section{\bf  Calculation of the viscosity coefficent}

We will calculate the shear viscosity coefficient by using the relativistic
kinetic theory for a massless QGP in the so-called relaxation time
approximation[2,3]:
$$\eta_i={4\over 15} \varepsilon_i\lambda_i,\eqno(1)$$

 where $\varepsilon_i$ is the energy density of particles of type $i$
 in the QGP  and $\lambda_i$ is the mean free path, which is the inverse
interaction rate: $\lambda_i={1\over\Gamma_i}$.
Considering the large angle scattering dominate the transport processes,
we should multiply the interaction rate by a factor $\sin^2\theta$,

$$\Gamma_{trsp.}=\int \sin^2\theta \Gamma_i, \eqno(2)$$
which is the so-called transport interaction rate, and
$\theta$ is the scattering angle in the center of mass system.

There are two equivalent ways of computing the interaction rates either
using the matrix elements or from  imaginary parts of the quark
or gluon self-energies [3].
$$\Gamma_i=-{1\over 4|p|} Im Tr(\not p\Sigma)|_{p_0=E}.\eqno(3)$$

Let us consider first the quark self-energy shown in
fig.1, where we have included the screening effects and the damping rate effects
by using the effective gluon propagator[8]. For hard quark momentum $(<p_q>\sim T)$
it is sufficient to use the bare quark propagator and bare vertices.
We will calculate the imaginary part of the self-energy at finite temperature
and chemical  potential using the Thermo Field Dynamics. For the hard particles
one can show that the main contributions to the interaction rate comes from the
soft momentum transfer range,i.e.$q\sim gT$ in the weak coupling limit[3].

Using the usual Feynman rules of QCD to evaluate the diagram of Fig.1,
one can obtain:
$$Tr(\not p\Sigma)=-g^2 C_f\int {d^D q\over (2 \pi)^D}{Tr( \not p \gamma^\mu
(\not p+\not q)\gamma^\nu) \over (p+q)^2} \Delta_{\mu\nu}, \ \ D=4-\epsilon ,\eqno(4)$$
where $\Delta_{\mu\nu}$ is the effective gluon propagator,
in covariant gauge[9],
$$\Delta_{\mu\nu}(q)=A_{\mu\nu}(q) \Delta_T(q)+B_{\mu\nu}\Delta_L(q)
+D_{\mu\nu}\Delta_\xi(q),\eqno(5)$$
 where
$$A_{\mu\nu}(q)=\delta_{\mu i}(\delta_{ij}-{q_i q_j\over Q^2})\delta_{j\nu},
\ \ \ B_{\mu\nu}=(\delta_{\mu 0}-{q_{\mu}q_0\over q^2}){q^2\over Q^2}
(\delta_{\nu 0}-{q_{\nu}q_0\over q^2})$$
$$D_{\mu\nu}={q_\mu q_\nu\over q^2},\ \ \
 \Delta_\xi=\xi{1\over q^2},
\ \ \ \Delta_{T,L}={1\over q^2-\Pi_{T,L}},\eqno(6)$$
where $\xi$ is the gauge parameter.

In the static limit the contribution of the hard thermal loops
for hot dense QCD are[5],
$$\Pi_L(q_0\to 0,Q)=\Pi_{00}(q_0\to 0,Q)=m_0^2
=g^2 T^2(1+{N_f\over 6}+{1\over 2\pi^2}\sum_f{\mu^2_f\over T^2}) \eqno(7)$$
 for the longitudinal part and
$$ \Pi_T(q_0\to 0,Q)=0 \eqno(8)$$
 for the transverse part of gluon polarization tensor.
Substituting eqs.(7),(8) into eq(6),one gets
$$\Delta_L={1\over q^2-m_0^2}, \ \ \ \ \Delta_T={1\over q^2}.\eqno(9)$$

In a thermal system, because of the thermal effects and the particles
interactions, the particles (quasi-particles) will no longer have
infinite lifetime, namely there are damping for the particles in QGP,
which is an important feature of QGP. If one take into account the
damping rate of the transverse gluon  $\nu_T\sim N \alpha_s T$[3,8],
the transverse
gluon propagator can be written as [6,7]
$$\Delta'_T={1\over (q_0-i\nu_T)^2-Q^2}.\eqno(10) $$
By using eqs(5)-(10),one can cast eq(4) into:
$$Tr(\not p\Sigma)=-g^2 C_f\int {d^D q\over (2 \pi)^D}
{2|P|^2\over (p+q)^2}
(\Delta'_T(q)-\Delta_L(q)-(1-\xi){(\hat P \cdot \hat Q)^2\over q^2})
.\eqno(11)$$
 where $C_f={N^2-1\over 2N}$.

For finite temperature QCD, we will use the hot dense propagators for
quarks and gluons in $2\times 2$ matrix
in TFD[9]. Among them the hot propagator of quark has the form

$$i\Delta_{11}(p)=-i\Delta_{22}(p)=\not p\bigl({1\over p^2-i\epsilon}-
2\pi(\theta(p_0) e^{x_p/2}n_f(x_p)+\theta(-p_0)e^{-x_p/2}n_f(-x_p))
 \delta(p^2)\bigr)$$

$$i\Delta_{12}(p)=-i \Delta_{21}(p)=-2\pi \not p e^{-\beta \mu}
\bigl(\theta(p_0) e^{x_p/2}n_f(x_p)+\theta(-p_0)e^{-x_p/2}n_f(-x_p)\bigr)
 \delta(p^2) \eqno(12)$$
where
$$n_f(x_p)={1\over e^{x_p}+1},\ \ \  \ x_p=\beta (p_0+\mu). $$

And for the gluons

$$\Delta_{11}^L(q)=-i\Delta_{22}^L(q)={1\over q^2-m_0^2-i\epsilon}-2\pi i\delta(q^2-m^2)n_B(q)
=\Delta_0^L+\Delta_\beta^L, $$
$$\Delta^{12}_L(q)=2\pi i\delta(q^2-m^2_0)n_B(q)e^{-\beta|q_0|/2}, $$
$$\Delta_{11}^T(q)= Re{1\over (q_0-i\nu_T)^2-q^2}
-2\pi i Im {1\over (q_0-i\nu_T)^2-q^2} n_B(q),$$
$$\Delta_{11}^T(q_0\to 0,q)= {1\over q^2-\nu_T^2-i\epsilon}
-2\pi i\delta(q^2-\nu^2)n_B(q),$$
$$\Delta^{12}_T(q)=2Im\Delta_T n_B(q)e^{-\beta|q_0|/2}$$
$$={1\over i|q_0|}[{1\over (q_0-i\nu_T)^2-Q^2}
-{1\over (q_0+i\nu_T)^2-Q^2}]n_B(q)e^{-\beta|q_0|/2}, \eqno(13)$$
where $n_B$ is the Bose distribution function and for soft mementum $q$
$$n_B(q)={1\over e^{\beta |q_0|}-1}\sim {T\over |q_0|},\ \ \ \
n_B(p+q)\sim n_B(p).$$

It is known that one can get the imginary part of the retard Green function
of the quark self-energy from the imaginary part of $(1-2)$ type Green
function,which is very convenient to calculate in TFD[10]
$$Im\bar\Sigma={e^{\beta\mu/2}\over \sin 2\phi_{p+\mu}} Im\Sigma_{12},
\eqno(14)$$

 where
$$\sin \phi_{p+\mu}={\theta(p_0) e^{-x_p/4}+\theta(-p_0)e^{x_p/4}
\over \sqrt{ e^{x_p/2}+e^{-x_p/2}}},  $$
$$\cos \phi_{p+\mu}={\theta(p_0) e^{x_p/4}+\theta(-p_0)e^{-x_p/4}
\over \sqrt{ e^{x_p/2}+e^{-x_p/2}}}. \eqno(15)$$

 Then one can obtain
$$Im Tr(\not p\Sigma)=4 g^2 C_f T \int {d^{D-1} q\over (2 \pi)^2}|p|^2
\delta ((p+q)^2)
({1\over q^2+\nu^2}-{1\over q^2+m^2}-(1-\xi){\cos^2\theta \over q^2}).
\eqno(16)$$

Substituting eq(16),into eq(2), and assuming the transfer momentum $q$ is soft:
$0<q\sim gT<T$, $q/P\sim g$ we obtain
$$\Gamma_{trsp,q}={g^2 C_f T\over 4\pi P^2}\int dq \int _{-1}^1 dz q^3
(1-z^2)\delta(z+{q\over 2P})({1\over q^2+\nu^2}-{1\over q^2+m^2}
-(1-\xi){z^2 \over q^2})$$
$$= {g^2 C_f T\over 4\pi P^2} \int dq  q^2
(1-O({q^2\over 4 P^2})) ({q^2\over q^2+\nu^2}-{q^2\over q^2+m^2}
-(1-\xi) O({q^2\over 4 P^2}))$$
which shows that the gauge dependent part is suppressed by 
the square of the coupling constant $g^2$ 
by evaluating the integration over $q$ in  $0<q\sim gT<T$, we obtain:
  $$\Gamma_{trsp,q}
={g^2 C_f T\over 8\pi P^2}(m^2 \ln {T^2+m^2\over m^2}+{\nu^2\over 2}
\ln{\nu^2\over T^2+\nu^2}). \eqno(17)$$

The  transport interaction rate of gluon can be obtained in a similar way
from the gluon self-energy,and it was found that
$$ \Gamma_{trsp,g}={N\over C_f}\Gamma_{trsp,q}.\eqno(18) $$

One can evaluate the  quark and gluon energy density
in QGP via the vaccum graphs in hot QCD both at finite
temperature and chemical potential:[1]
$$\varepsilon_g=16\times {\pi^2\over 30} T^4(1-{15\alpha_s\over 4\pi}) $$
$$\varepsilon_q=6N_f( {7\pi^2\over 120} T^4 (1-{50\alpha_s\over 21\pi}) +{1\over 4}\mu^2 T^2
 +{\mu^4\over 8\pi^2}(1-{2\alpha_s\over 2\pi})) .\eqno(19)$$

By making use  of eqs(1),(2),(19), and assuming $p\simeq 3T$,
$\nu_T\sim N \alpha_s T$[8]
we can get the shear viscosity coefficient of quarks for two flavors and gluons
respectively
$$\eta_q
={{27\over 5\pi}( {7\pi^2\over 120} T^3 (1-{50\alpha_s\over 21\pi}) +{1\over 4}\mu^2 T
 +{\mu^4\over 8\pi^2 T}(1-{2\alpha_s\over 2\pi}))
 \over \alpha_s^2(4/3+{1\over \pi^2}{\mu^2\over T^2})
(-\ln \alpha_s+\ln ( {1\over 4\pi(4/3+{1\over \pi^2}{\mu^2\over T^2})}+\alpha_s)
+A(\nu))} . \eqno(20)$$
$$\eta_g={{16\over 5\pi}
( {\pi^2\over 30}) T^3(1-{15\alpha_s\over 4\pi})
\over \alpha_s^2(4/3+{1\over \pi^2}{\mu^2\over T^2})
(-\ln \alpha_s+\ln ( {1\over 4\pi (4/3+{1\over \pi^2}{\mu^2\over T^2})}+\alpha_s)
+A(\nu))} .\eqno(21)$$
$$A(\nu)={\nu^2\over m^2}\ln{\nu^2\over \nu^2+T^2}\sim
{N_c^2\alpha_s \over 4\pi ( 4/3+{1\over \pi^2}{\mu^2\over T^2})}
\ln( \alpha_s^2 {N_c^2\over 1+N_c^2 \alpha_s^2}).\eqno(22) $$

In order to compare with previous calculations,one may
 come back to vanishing quark chemical potential case and not
take the gluon damping rate into account, namely  substituting $A(\nu)=0$
in eqs(20)-(22),then
one obtains the shear
viscosity coefficient of QGP with two quark flavors at vanishing
quark chemical potential without including the finite width effect:
$$\eta={T^3 \over \alpha_s^2}[ {0.25(1-{50\alpha_s\over 21\pi})\over \ln({0.6 +\alpha_s\over \alpha_s})}
+ {0.73(1-{15\alpha_s\over 4\pi})\over \ln({0.6+\alpha_s\over \alpha_s})}],\eqno(23)$$
 which agrees with the previous result in ref.3 beyond the leading logarithm
approximation in the weak coupling limit
 except a factor $2$ due to
the definition. At leading order and leading logarithm  approximation,
one can cast eq(23)
into
$$\eta=0.98{T^3\over \alpha_s^2 \ln{ 1/ \alpha_s}}.\eqno(24)$$
 The coefficient $ c=0.98$ is
remarkably close to the one found by Thoma  $c=1.02$ [3] and the one found by
Baym et al.[2], the reason of the difference is we used $<p_q>\sim<p_g>\sim 3T$
other than $<p_q>\sim 3.2T$, $<p_g>\sim 2.7T$.

\section{\bf Numerical analysis and discussion}

Before our numerical analysis, we consider the running behavior of the
coupling constant g. In principle the running coupleing constant is governed
by the renormalarization equations of QCD at finite temperture and chemical 
potential , which is still under investigation . Hereby we only use the
following result[12]:

$$\alpha(T,\mu)={12\pi \over (33-2 n_f)\ln{0.8 \mu^2 
+15.622 T^2\over \Lambda_s^2}}, \eqno(25)$$
where 
is the QCD papameter, and we choose $\Lambda_s=0.1GeV$ as usual.
And $n_f$ is the number of flavor, we will 
only consider two flavors.   
 
Substituting the running coupling constant eq(25) into egs(20)-(22),
we obtain the experessions of the shear viscosity coefficients as 
functions of temperature and chemical potential. So we can plot the 
curves of the quark and gluon viscosity coefficients against 
 temperature in Fig.2(a) and Fig. 3(a)
 for different  quark chemical potentials $\mu=0,0.2,0.4 GeV$ 
with and without including the gluon damping
effect in QGP respectively.
 From Fig.2(a) one can find the quark chemical potential effect
may enhance
the  quark viscosity coefficient tremendously because the increase of
quark energy density in QGP, while  the chemical potential effect on the gluon
viscosity is almost negligible.  From Fig. 3(a)
we find that the three curves of gluon viscosity coefficient against $T$ at different chemical potential
almost meet together both with and without taking into acount the gluon finite
width.
   On the other hand the gluon damping effect causes
large enhancements to the shear viscosities both  for  quark and gluon.
The reason
is that the damping effect has decreased their transport interaction rates,and hence
increase their mean free paths.This qualitatively agrees with the analyses in
ref.4 for the nuclear matter.

Fig.2(b) 
and Fig.3(b) show the temperature dependence of the scaled 
viscosity coefficients
$\eta/T^3$   for different  quark chemical potentials $\mu=0,0.2,0.4 GeV$ 
with and without including the gluon damping effect . 
Our result  agrees with the Kubo formula result
in ref.2
$\eta/T^3\ge 2.6/\alpha_s$ at vanish baryon density, but our result is much bigger than the lattice
estimate near the critical temperature in ref.2 . Fig.2(b) and Fig.3(b) showes
at vanish chemical potential the curves of scaled viscoisity coefficients  are no longer flat lines because of the running coupling constant.
On the other hand, the tendencies  are quite different between the scaled 
viscoisity of quark and gluon. The reason is the quark chemical potential 
plays different roles in them. Neverthless, the finite width still cause
considerable enhancement to both the scaled viscosity coefficients

Now we study how the shear viscosity coefficients vary with the
coupling constant. We plot the curves of the 
scaled viscosity coefficients
$\eta/T^3$  against the coupling constant   at  a fixed temperature $T=300MeV$
and different  quark chemical potentials $\mu=0,0.2,0.4 GeV$ 
with and without including the gluon damping effect  in Fig.2(c) and Fig.3(c). 
The curves are almost above the Navier-Stockes Regime and in agreement with
Thoma's result. Also the tendencies  are quite different between the scaled 
viscoisity of quark and gluon via the coupling constant because of
  quark chemical potential 
plays different roles in them. However , the finite width still cause
considerable enhancement to both the scaled viscosity coefficients.
One interesting thing is that, although  our calculations are valid for weak
coupling limit $g<1$, Fig.2(c) and Fig.3(c)
 show that the results can be extrapolated to higher coupling constant
at fixed  temperatur $T=300MeV$, as predicted in ref.3 at least for
quantities of energetic particles with $E>>T$.

\section{\bf  Summary and Conclusions}

We have examined the quark and gluon shear viscosity behaviors by real time QCD  both
at finite temperature and baryon density in the quark gluon plasma in which
both Debye screening and damping or finite width are regarded as important
underlying features. Also the running feature of the coupling constant is accounted for. We have seen that both the quark and gluon shear viscosity 
coefficients are increased substantially,due to the effects of the finite width
of transverse gluons. Meanwhile all the scaled shear viscosity coefficients
are increased in comparison with the previors results,in which the finite widtheffect was not accounted for.

We have also examined the quark chemical potential effects to the
shear viscosity coefficients in a straight forward way. It also enhances
the shear viscosity coefficients of QGP consiberably, but the enhancements are smaller than
that caused by the finite width effect. On the other hand, we have seen
the tendencies of the quark viscosity coefficients, especially the scaled ones,differ quite substantially from that of the gluon. This is because the
baryon density plays different roles  in quark and gluon shear 
viscosity coefficients.

We note that If we remove the quark chemical potential and finite width
the results will be in  good agreement with
 previous calculations [2,3,13].

We conclude that both
the gluon damping rate and  quark chemical potential effects have a 
considerable influence on the dissipative
processes of QGP. This dissipation cannot be neglected in hydrodynamic descriptions
of the expansion phase of QGP in ultrarelativistic heavy ion collisions.

\section*{\bf Acknowledgments}

This work is supported partly by the National Natural Science Funds of China.
Hou Defu wants to thank the Deutsche Forschungsgemeinschaft (DFG) for supports.
The authors are indebted to Peter Henning and Ulrich Heinz for valuble discussion and comments,and  whish to 
thank  Liu Lianshou for his help.

\newpage
\section*{Reference}
\begin{description}

\item[[1]]
 M\"uller, in NATO ASI Series B, Physics Vol.303(Plenum Press, NewYork, London,1992);
 U.Heinz, K.Kajantie and T.Toimela, Annals of Phys.{\bf 176},218(1987);
   J.-P. Blaizot, Nucl. Phys. {\bf A566}, (1994)333c.

\item[[2]]
 A.Hosoya and K.K. Kajantie, Nucl.Phys.{\bf B250},666(1985);
   P. Danielewicz and M.Gyulassy, Phys.Rev.{\bf D31},53(1985);

G.Baym, H. Monien, C.J.Pethick and D.G.Ravenha, Phys.Rev.Lett.{\bf 64}
   , 1867(1990).

\item[[3]]
 M.H.Thoma , Phys.Rev.{\bf D49}(1994)451;
   M.H.Thoma , Phys.Lett{\bf B}(1991)144.

\item[[4]]
 P.A.Henning, Nucl.Phys.{\bf A567},844(1994).

\item[[5]]
 H. Vija and  M.H.Thoma Phys.Lett{\bf B342},212(1995).

\item[[6]]
 N.P.Landsman, and Ch.G.Van Weert, Phys.Rep. {\bf 145},141(1987).

\item[[7]]
 P.Henning Phys.Rep.{\bf 253},235 (1995);
 Hou Defu, Li Jiarong, Z.Phys.{\bf C71},503(1996)
 Hou Defu, S. Ochs, Li Jiarong, Phys. Rev.{\bf D54},(1996)(in press)

\item[[8]]
 R.D.Pisarski, Phys.Rev.Lett.
   {\bf 63},1129(1989);
   E.Braaten, R.D.Pisarski, Phys. Rev.{\bf D42}, 2156(1990), Phys.Rev.Lett.,
   {\bf 64},1338(1990).

\item[[9]]
  R.Kobes, Phys. Rev. {\bf D42},562 (1990);
     Phys. Rev. {\bf D43},1269(1991).

\item[[10]]
 Y.Fujimoto, M. Morikawa and M. Sasaki, Phys. Rev. {\bf D33},590(1986).

\item[[11]]
  Y. Fujimoto and Hou Defu, Phys. Lett. {\bf B335}, (1994)87.

\item[[12]]
C.P.Singh B.K.Patra, and Saeed Uddin, Phys.Rev. {\bf D49},4023(1994).

\item[[13]]
  P.Rehberg, S.P. Klevansky and J.H\"ufner, Nucl.Phys.{A608 },356(1996).

\end{description}

\newpage
\section*{Figure captions}

\begin{description}

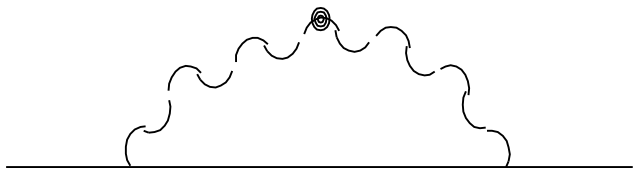
\item[Fig.1]
{The resummed one-loop quark self-energy graph }\label{fig.1}


\item[Fig.2(a)]
{ The quark viscosity coefficient as a function of the temperture for

$\mu=0,0.2,0.4$  (bottom to top) without (thin solid lines in lower part )
and with (thick solid lines in upper part)including the gluon damping effect .}\label{fig.2(a)}

\item[Fig.2(b)]
{ The scaled quark viscosity coefficient $\eta/T^3$ as a function of the temperture for
$\mu=0,0.2,0.4$  (bottom to top) without (thin solid lines in lower part )
and with (thick solid lines in upper part)including the gluon damping effect .}\label{fig.2(b)}
 
\item[Fig.2(c)]
{ The scaled quark viscosity coefficient $\eta/T^3$ as a function of 
coupling constant $\alpha$ at fixed temperature $T=300MeV$  for
$\mu=0,0.2,0.4$  (bottom to top) without (thin solid lines in lower part )
and with (thick solid lines in upper part)including the gluon damping effect .}\label{fig.2(c)}

\item[Fig.3(a)]
{ The gluon viscosity coefficient as a function of the temperture for
$\mu=0,0.2,0.4$  (meet together) without (thin solid lines in lower part )
and with (thick solid lines in upper part)including the gluon damping effect .}\label{fig.3(a)}

\item[Fig.3(b)]
{ The scaled gluon viscosity coefficient $\eta/T^3$ as a function of the temperture for
$\mu=0,0.2,0.4$  (bottom to top) without (thin solid lines in lower part )
and with (thick solid lines in upper part)including the gluon damping effect .}\label{fig.3(b)}

\item[Fig.3(c)]
{ The scaled gluon  viscosity coefficient $\eta/T^3$ as a function of 
coupling constant $\alpha$ at fixed temperature $T=300MeV$  for
$\mu=0,0.2,0.4$  (bottom to top) without (thin solid lines in lower part )
and with (thick solid lines in upper part)including the gluon damping effect .}\label{fig.3(c)} 

\end{description}


\end{document}